\documentclass[a4paper]{jpconf}
\usepackage{graphicx}
\usepackage{amssymb,amsmath}

\begin{document}
\title{Symmetric coupling of angular momenta, quadratic algebras and discrete
 polynomials\,\footnote{\,{\em J. Phys.: Conf. Ser.} (2014), in press.}}

\author{V Aquilanti$^{\dag}$, D Marinelli$^{\ddag}$ and A Marzuoli$^{\S}$ }

\address{$^{\dag}$ Dipartimento di Chimica, Biologia e Biotecnologie, via Elce di Sotto 8, Universit\`a  di Perugia,  06123 Perugia, I}

\address{$^{\ddag}$ Dipartimento di Fisica, Universit\`a di Pavia \& INFN Sezione di Pavia, via Bassi 6, 
27100 Pavia, I}

\address{$^{\S}$ Dipartimento di Matematica `F Casorati', Universit\`a di Pavia, via Ferrata 1 \& INFN Sezione di Pavia, via Bassi 6, 27100 Pavia, I}

\ead{$^{\dag}$vincenzoaquilanti@yahoo.it, $^{\ddag}$dimitri.marinelli@pv.infn.it, $^{\S}$annalisa.marzuoli@pv.infn.it}

\begin{abstract}
Eigenvalues and eigenfunctions of the volume operator, associated with the symmetric coupling of three $SU(2)$ angular momentum operators, can be analyzed on the basis of
a discrete Schr\"odinger--like equation which provides a  semiclassical Hamiltonian picture 
of the evolution of a `quantum of space', as shown by the authors in \cite{AqMaMaJPA2013}.
Emphasis is given here to the formalization in terms of a quadratic symmetry algebra
and its automorphism group. This view is related to the Askey scheme, the  hierarchical structure which includes all hypergeometric polynomials of one (discrete or continuous) variable. Key tool for this comparative analysis is the duality operation defined on the generators of the quadratic algebra and suitably extended to the various  families of overlap functions (generalized recoupling coefficients). These families, recognized as lying at the top level of the Askey scheme, are classified and a few limiting cases are addressed.
\end{abstract}

\section{Introduction and a brief review}\label{intro}
In  \cite{AqMaMaJPA2013} a family of orthogonal polynomials has been introduced based on a three--term recursion relationship which plays the role of a discrete Schr\"odinger equation describing the action of a `volume' operator. This operator occurs in the symmetric treatment of the quantum few--body problem as well as in spin--network modeling  of a quantum of space, as pointed out originally in  \cite{CaCaMaCQG2002}. In this section a short introduction to the necessary mathematical background will be given, together with a summary of a few significant results found by the authors  in \cite{AqMaMaJPA2013}. 
Improved insights into algebraic and analytical aspects of the subject will be provided in the next sections. \\

The theory of (re)coupling of eigenstates of three $SU(2)$ angular momentum operators 
${\bf J}_1$, ${\bf J}_2$, ${\bf J}_3$ to states of sharp total angular momentum 
${\bf J}_4$ (with projection $J_0$ along
the quantization axis) is  usually carried out  in the setting of  `binary couplings'
(see  \cite{BiLo9}, Topic 12 and original references therein). 
Referring to the ordered triple as above,
 the admissible schemes are ${\bf J}_1+{\bf J}_2={\bf J}_{12};\;{\bf J}_{12}+{\bf J}_3=
{\bf J}$ and 
${\bf J}_2+{\bf J}_3={\bf J}_{23};\;{\bf J}_1+{\bf J}_{23}={\bf J}_4$, respectively.
The corresponding eigenvectors are denoted
\begin{equation}\label{jbasis}
|j_{12}\,> \,:= \,|\,(j_1j_2)j_{12}\,j_3\,j_4m>\; \text{and}\;
|j_{23}\,> \, := \,|\,j_1(\,j_2j_3)j_{23}\,j_4m'>\,,
\end{equation}
where small $j$s are labelings of  the eigenvalues associated with the angular momentum
operators ({\em e.g.} ${\bf J}_1^2|j_{12}\,>=j_1(j_1+1)\,|j_{12}\,>$ , ${\bf J}_{12}^2|j_{12}\,>$ $=j_{12}(j_{12}+1)\,|j_{12}\,>$) running  over $\{0, 1/2, 1, 3/2, \dots\}$, in $\hbar$ units,
and $m$ ($m'$) is the eigenvalue of $J_0$ with $-j_4 \leq m,m' \leq j_4$ in integer steps.
Thus the ket vectors above belong to Hilbert spaces representing  
simultaneous eigenspaces of the two, partially overlapping sets of 
commuting operators ${\bf J}_1^2$, ${\bf J}_2^2$, ${\bf J}_3^2$, ${\bf J}_4^2$, $J_0$ and
${\bf J}_{12}^2$ (${\bf J}_{23}^2$, respectively). 
The Racah--Wigner $6j$ symbol is  the unitary (actually orthogonal by Condon-–Shortley convention) 
transformation relating the two sets  (\ref{jbasis}) according to  
\begin{equation}\label{seij}
<j_{23}\,|\,j_{12}\,>\;:=
\,(-1)^{\Phi}\,
 [(2j_{12}+1)\,(2j_{23}+1)]^{1/2}\,
\,\begin{Bmatrix}
j_1 & j_2 & j_{12} \\
j_3 & j_4 & j_{23}
\end{Bmatrix},
\end{equation}
where $\Phi \equiv j_1+j_2+j_3$ and the weights $(2j+1)$ are the dimensions of the spin--$j$ representations of $SU(2)$ which provide the standard normalization of such a `recoupling coefficient' as encoded in the shorthand notation in the left--hand side. Therefore a basis transform is simply written as $|j_{23}\,>\, = \sum_{j_{12}'}$
$<j_{23}\,|\,j_{12}'\,>$ $\,|j_{12}'\,>$ while the inverse one is achieved by the transpose $<j_{12}\,|\,j_{23}\,>$
(all non--null matrix elements obey the selection rule  $m=m'$ by Wigner--Eckart theorem). Recall in passing that the $6j$ symbol in (\ref{seij}) encodes naturally the symmetry of an Euclidean tetrahedron, a fact which is at the basis of the huge amount of literature about `spin--network' models for 3--dimensional discretized quantum gravity and quantum computing flourished in the past two decades (see \cite{CaMaRa2009} and references therein). 

The treatment of the `symmetric' coupling scheme
for the addition of three $SU(2)$ angular momenta $\mathbf{J}_1,\,\mathbf{J}_2, \,\mathbf{J}_3$ to give  $\mathbf{J}_4$ (with projection $J_0$) is characterized in terms of a `volume' operator $K=\mathbf{J}_1\cdot$ $\mathbf{J}_2\times\mathbf{J}_3$. Unlike what happens with  binary coupling schemes, the $\mathbf{J}$s 
appear now to be all on the same footing, indicating that the volume operator can be thought of as acting  democratically on either a composite system of four objects with vanishing total angular momentum (a configuration that can be associated with a not necessarily planar quadrilateral vector diagram ${\bf J_1 +J_2 + J_3 +J_4}=0$), or a system of three objects with total angular momentum $\mathbf{J}_4$ (see again \cite{BiLo9}, Topic 12, last section, and original references therein). The present scheme is  characterized by  the six commuting Hermitian operators ${\bf J}_1^2$, ${\bf J}_2^2$, ${\bf J}_3^2$, $K$, ${\bf J}^2$ and $J_0$, so that eigenvectors and eigenvalues  of $K$ are given formally 
(consistently with the notation used in  (\ref{jbasis})) as
\begin{equation}\label{kbasisl}
|k\,> \, :=\,|\,(j_1j_2j_3)\,k\,j_4 m> \;\; \text{with}\;\; K\,|k\,>\,=\,\lambda_k\,|k\,>\,.
\end{equation}
Eigenvalues and matrix elements of $K$ are naturally found within  an imaginary antisymmetric representation based on a three--terms recursion relationship  \cite{CaCaMaCQG2002}, which can be turned   into a  real, time--independent Schr\"odinger  equation which governs  the  dynamics of a `quantum of space' as a function of a discrete variable denoted $\ell$ (see below). This has been achieved in \cite{AqMaMaJPA2013} (which can be referred to also for a complete list of  previous and related  papers) where the introduction of discrete, potential--like functions highlights the surprising  crucial role  of `hidden' symmetries, first discovered by Regge \cite{Regge59} for the  $6j$ symbols. 
The  Schr\"odinger equation is discretized with respect to a lattice variable given by the label of the operator ${\bf J}_{12}$ which characterizes the first of the binary schemes in (\ref{jbasis}) and reads
\begin{equation}\label{SymRR}
\lambda_k\,\Psi_{\,\ell}^{(k)}\,\,+\alpha_{\,\ell+1}\;\Psi_{\,\ell+1}^{(k)}\,+\alpha_{\,\ell}\;
\Psi_{\,\ell-1}^{(k)}=0\;\; \text{with} \;\; \;\ell \equiv j_{12} \,  \in 
\{j_{12}^{\text{min}},\,j_{12}^{\text{min}}+1,\,
\dots , j_{12}^{\text{max}} \}\,,
\end{equation}
where the matrix elements $\alpha_\ell$  are expressed in terms of geometric quantities, namely
\begin{equation}\label{alphal}
\alpha_{\ell}=F(\ell;j_{1}+1/2;j_2+1/2)F(\ell;j_3+1/2;j_4+1/2)\,/\,[(2\ell+1)(2\ell-1)]^{1/2}\,.
\end{equation}
Here $F(A,B,C))  =\tfrac{1}{4} [(A+B+C)  
\ (-A+B+C)\ (A-B+C)\ (A+B-C)] ^{\tfrac{1}{2}}$ is the Heron's formula for the  area of a triangle with side lengths $A,B$ and $C$.  Thus   $\alpha_\ell$ is  proportional to the product of the areas of  the two triangles sharing the side of length $\ell$ and forming a quadrilateral of sides $j_1+\tfrac{1}{2},\, j_2+\tfrac{1}{2},\, j_3+\tfrac{1}{2}$ and $j_4+\tfrac{1}{2}$. Such a parameter quadrilateral, together with its Regge--conjugate
(see below), is the guiding tool of the combinatorial and geometric analysis, in the asymptotic limit, of the  Hamiltonian 
dynamics governing both tetrahedral and `fluttery' quadrilateral configurations, see sections 2 and 3 of \cite{AqMaMaJPA2013} for more details. 


\section{Quadratic symmetry algebras}\label{algebra}

Following \cite{GrLuZhAOP1992,GrZhSovPh1988}, 
the quantum version of a classical dynamical algebra associated with
a pair of `mutually integrable' dynamical variables calls into play a triple 
$K_1,K_2,K_3$ of linear operators acting on a (suitably defined) Hilbert space 
with $K_{1,2}$ Hermitian and algebraically independent and $K_3 :=[K_1,K_2]$ anti--Hermitian. 
The request that these generators do fulfill the Jacobi identity constrains 
the fundamental commutation relations 
to be of the form ($\{\,,\,\}$ is the anticommutator)
\begin{equation}\label{CommRel}
\begin{split}
  [K_1,K_2] &= \, K_3 \\
[K_2,K_3] &= \, 2R\,K_2K_1K_2 +A_1\, \{ K_1,K_2\} +A_2\,K_2^2 + C_1\,K_1 + D\,K_2 + G_1\\
[K_3,K_1] & =\, 2R\,K_1K_2 K_1 + A_1\, K_1^2 + A_2\,\{ K_1,K_2\} + C_2\,K_2 + D\,K_1 + G_2\,,
\end{split}
\end{equation}
where $R, A_{1,2}, C_{1,2}, D, G_{1,2}$ are real parameters (the structure constants) and the right--hand sides of the last two relations contain only Hermitian terms. Such a kind of algebraic structures was actually introduced by Sklyanin \cite{Skly1982} and they are called  `quadratic' algebras for the obvious reason that the commutators (Poisson brackets in the classical cases) are  combinations of quadratic (and linear) terms in each of the generators. Mutual 
integrability is a sharper requirement with respect to the original formulation, and  amounts to look at the symmetry algebra as a dynamical one --where $K_1$ is a constant of the motion for $K_2$ taken as the Hamiltonian operator, as well as the other way around.
Further improvements in the the study of classical, quantum and $q$-deformed symmetries along these lines have been provided over the past few decades by a number of authors. Often the admissible structures associated with \eqref{CommRel}
and listed in the table below \cite{GrLuZhAOP1992} are referred to as `Zhedanov's algebras' in the literature. 
Note that for completeness the last line includes the two `standard' Lie algebras on three generators (whose commutation relations are by definition linear).\\

\begin{center}
{\bf Classification of quadratic algebras}
\end{center}
\begin{center}
\begin{tabular}{||l||c|c|c|c||}
\, & $R$ & $A_1$ & $A_2$ & $C\, \& \,D$ \\ \hline \hline
{\bf AW(3)} ({\em Askey--Wilson}) & * & * & * & * \\ \hline
{\bf R(3)} ({\em Racah}) & 0 & * & * & * \\ \hline
{\bf H(3)} ({\em Hahn}) & 0 & 0 & * & * \\ \hline
{\bf J(3)} ({\em Jacobi}) & 0 & 0 & * & 0 \\ \hline
{\bf Lie} algebras:  & 0 & 0 & 0 & * \\
$su(2)$, $su(1,1)$ & \, & \, & \, & \, \\ \hline \hline
\end{tabular}
\end{center}

The denominations of the algebras, Askey--Wilson, Racah, ..., are strongly reminiscent of the Askey--Wilson scheme of hypergeometric orthogonal polynomials of one (continuous or discrete) variable \cite{Askey}. This is not  accidental:
 rather, this remark turns out to be  crucial  in order to recognize the deep connection between  algebraic symmetries of (quantum) systems and special function theory in a quite straightforward way. 
Indeed the `overlap functions' stemming from the analysis of the eigenvalue problems for the operators $K_1,K_2,K_3$ which generate the quadratic algebras are, under mild conditions, orthogonal families of  Wilson, Racah,  Hanh, Jacobi, ..., Hermite polynomials. In what follows an account of a few technical details is given for the case of the Racah algebra 
{\bf R(3)} which corresponds to set $R=0$ in (\ref{CommRel}).\\

Suppose that the Hermitian operators $K_1$ and $K_2$ --defined on a separable Hilbert space and possibly depending on a same (finite) set of real parameters-- are both ladder operators, namely possess discrete, non--degenerate spectra, and start considering the eigenvalue problem for $K_1$
\begin{equation}\label{K1spec}
 K_1\, \psi_p \,=\, \chi_p \,\psi_p\,, \;\;\;  p=0,1,2,\dots \,.
 \end{equation}
Then it can be easily shown that the operator $K_2$ is tridiagonal in this basis
\begin{equation}\label{K23term}
K_2\, \psi_{p} \,=\, 
\mathfrak{a}_{p+1} \,\psi_{p+1} \,+ \, \mathfrak{a}_p \,\psi_{p-1} \,+ \, \mathfrak{b}_p \,\psi_p 
 \end{equation}
and, similarly, by exchanging the role of $K_1$ and $K_2$, one would get
\begin{equation}\label{K2spec}
  K_2\, \phi_s \,=\, \mu_s \,\phi_s , \;\;\; s=0,1,2,\dots 
 \end{equation}
and
 \begin{equation}\label{K13term}
  K_1\, \phi_s \,=\, \mathfrak{c}_{s+1} \,\phi_{s+1} \,+ \, 
\mathfrak{c}_s \,\phi_{s-1} \,+ \, \mathfrak{d}_s \,\phi_s\,.
 \end{equation}
The (real) matrix coefficients $\mathfrak{a}, \mathfrak{b},$ $\mathfrak{c}, \mathfrak{d}$ can be evaluated explicitly in terms of the  commutation relations  \eqref{CommRel} and contain also the parameters which the operators may depend on
(such parameters are dropped in the present simplified treatment aimed to point out the overall  structural properties). Once chosen suitable normalizations for the two sets of eigenbases (\ref{K1spec}) and  (\ref{K2spec}), 
it is possible to introduce two families of overlap functions
by resorting to the Dirac braket convention 
(in which for instance  $<x|\psi>$ stands for the eigenfunction  $|\psi\,>$ of a system in the position representation) 
 \begin{equation}\label{over1e2} 
<\phi_s\,|\psi_p\,> \,\equiv \, <s\,|\psi_p\,> \,\equiv \, <s\,|p\,>\;\;\text{and}\;\;<\psi_p\,|\phi_s\,> \,\equiv \, <p\,|\phi_s\,>\,\equiv \, <p\,|s\,>
 \end{equation}
which are both hypergeometric orthogonal polynomials of one discrete variable (the spectral parameter $\mu_s$ and 
$\chi_p$ respectively) to be identified, up to suitable rearrangements of the hidden parameters, with the Racah polynomial on the top of the Askey scheme \cite{Askey}.

In the $K_1$--eigenbasis  the operator $K_3$ satisfies
 \begin{equation}\label{K33term}
 K_3\, \psi_p \,=\, (\chi_{p+1}-\chi_{p})\, \mathfrak{a}_{p+1} \,\psi_{p+1} \,- \, (\chi_{p}-\chi_{p-1})\,\mathfrak{a}_p \,
\psi_{p-1}\,, 
 \end{equation}
where eigenvalues $\chi$ and  matrix elements $\mathfrak{a}$ are iteratively evaluated from (\ref{K1spec}) and (\ref{K23term}).
$K_3$ has a discrete spectrum found as a solution of   
\begin{equation}\label{K3spec}
 K_3 \, \varphi_n \,=\, \nu_n \,\varphi_n , \;\;\; n=0,1,2,\dots\;.
 \end{equation}
It is worth noting that in general the diagonalization  of $K_3$ cannot be carried out analytically, except in a few  cases in which at least the lowest eigenvalues turn out to be representable in closed algebraic forms. The  associated families of (normalized) overlap functions are denoted
\begin{equation}\label{over1e3} 
<\varphi_n\,|\psi_p\,> \,\equiv \,<n|p>\;\;\text{and}\;\;
<\psi_p\,|\varphi_n\,>\,\equiv \,<p|n>\;\;\; (n=0,1,2,\dots ; p=0,1,2,\dots)
\end{equation}
and can be shown to be orthogonal (on different suitably defined lattices), each depending on one discrete variable, but in principle they might not be included  into the Askey scheme.\\
Similarly, other two families of (normalized) overlap functions associated with the pair $K_2, K_3$ can be defined by notation consistency as 
\begin{equation}\label{over2e3} 
<\varphi_n\,|\phi_s\,> \,\equiv \,<n|s>\;\;\text{and}\;\;
<\phi_s\,|\varphi_n\,>\,\equiv \,<s|n>\;\;\; (n=0,1,2,\dots ; s=0,1,2,\dots)\,.
\end{equation}

A crucial feature of the Racah algebra {\bf R(3)} and associated overlap functions is the duality property. 
It relies on the following transformation of the generators  
\begin{equation}\label{ExGen} 
K_1 \leftrightarrows K_2\,\,; \;\;\;K_3 \rightarrow -K_3
\end{equation}
which can be easily shown to represent an automorphism of the Racah algebra {\bf R(3)}. The notion of duality
is extended to (all of) the sets of overlap functions introduced above. More precisely 
\begin{itemize}
\item[{\bf i)}] Under the automorphism (\ref{ExGen}) the discrete variables of the two hypergeometric families of overlap functions associated with $K_1, K_2$ given in (\ref{over1e2}) and their degrees as  polynomials are interchanged. Since in the present case the operator $K_3$ is not called into play,
the stronger property of `self--duality' of these  families  holds true: both of them are recognized as Racah polynomials, as already mentioned above.   
\item[{\bf ii)}] Referring to the families in (\ref{over1e3}), under the automorphism (\ref{ExGen}) the discrete spectral variable  $\nu_n$ of the first family, which is orthogonal on the  lattice $p=0,1,2,\dots$, is turned into the second family, where the variable is $p$ and the polynomial degree is given in terms of  the labels $n=0,1,2,\dots$ of the eigenvalues of $K_3$.\\
A similar property is shared by the families associated with the pair $K_2, K_3$ given in (\ref{over2e3}).
\end{itemize}
More details on the nature of the automorphism group and on the statements about the overlap functions  will be reported  in the next section when dealing with a specific `realization' of the Racah algebra. 


 \section{Generalized recoupling theory, Regge symmetry and duality}\label{Regge}
 
The realization of the Racah algebra {\bf R(3)}  within the setting of generalized $SU(2)$ recoupling theory was actually the issue addressed originally in \cite{GrZhSovPh1988} which  has inspired further work on quadratic algebras. Combining the definitions and notation of section
\ref{algebra} with those of section \ref{intro} it is straightforward to recognize the following correspondence
\begin{align}\label{dict1}
K_1 & =\,\mathbf{J}_{12}^2\,;\;\;
K_2 \,=\,\mathbf{J}_{23}^2\,; \nonumber \\
K_3 & =\, [\mathbf{J}_{12}^2,\mathbf{J}_{23}^2]\,=\,-4i\, \mathbf{J}_1 \cdot (\mathbf{J}_2 \times \mathbf{J_3})\,\equiv -4i\,K
\end{align}
between the abstract ordered set of operators $K_1, K_2, K_3$ and its realization as $\mathbf{J}_{12}^2,$
$\mathbf{J}_{23}^2\,, K$.\\
The next step would consist in associating eigenvalue equations  and three--term recursion relations of the abstract approach with their realizations in generalized quantum (re)coupling theory. Here we do not enter into much details about this matter since the translation of (\ref{K23term}) based on the pair $K_1$, $K_2$ represents the three--term recursion relation for the $6j$ coefficient in disguise (see {\em e.g.} \cite{Russi}). The analysis for the pair  $K_1$, $K_3$ which gives the abstract three--term relation as written in 
(\ref{K33term}) is examined in details in \cite{CaCaMaCQG2002} (and references therein) while its symmetrized counterpart is nothing but the discretized Schr\"odinger--like equation displayed already in  (\ref{SymRR}) \cite{AqMaMaJPA2013}.

Focusing on the specific issue regarding the families of solutions of such relationships, one would directly be lead to establish the correspondence
\begin{equation}\label{dict2} 
\text{overlap functions}\;\;\longrightarrow \;\; 
\text{binary and symmetric recoupling coefficients}\,,
\end{equation}
where the arrow stands for the specific realization (\ref{dict1}) of {\bf R(3)}.
To achieve this goal in a transparent and consistent  way a few more steps are needed, the first one of which consists in establishing suitable notations for all of the recoupling coefficients. 
The $6j$ symbol in (\ref{seij}) and the functions $\Psi_{\,\ell}^{(k)}$ in 
(\ref{SymRR}) are thus  denoted and defined respectively  as 
\begin{equation}\label{Conv1}
<j_{23}\,|\,j_{12}\,>\,\equiv <\tilde{\ell}\,|\,\ell\,> \;\;\text{and} \;\;
\,\Psi_{\,\ell}^{(k)} \, :=\, <\ell\,|\,k\,>\,.
\end{equation}
Actually this is not  a mere question of notation, since in this way the objects 
 $<\bullet|\circ>$  may reveal their 
 `double' meaning as {\em i)} quantum mechanical transition amplitudes, 
namely the square modulus
 $|<\bullet|\circ>\,|^2$ is the probability that a system, prepared in the state  $|\circ>$, be measured to be in the state $|\bullet>\,$; {\em ii)} eigenfunctions of the operator whose quantum number is in
$|\circ>$ in the representation labeled by the eigenvalue of the other operator, namely through
the projection  onto $<\bullet|\,$. The latter interpretation 
will be under focus in what follows and more details about the correspondence (\ref{dict2}) can be worked out by introducing explicitly 
the (so far ignored) parameters of the problem. Upon replacement of the original (ordered) set of labeling of the four angular momenta forming a quadrilateral according to
\begin{equation}\label{Conv2}
(\,j_1, j_2, j_3, j_4\,) \;\;\mapsto\;\;(\,a,b,c,d\,)\,,  
\end{equation}
the functionals are rewritten as
\begin{equation}\label{Conv3}
 <\tilde{\ell}\,|\,\ell\,>\,(a,b,c,d) \;\propto
\begin{Bmatrix}
a & b & \ell \\
c & d & \tilde{\ell}
\end{Bmatrix}\
\;\;\text{and} \;\;
\,\Psi_{\,\ell}^{(k)}(a,b,c,d) \,=\, <\ell\,|\,k\,>\,(a,b,c,d)\,. 
\end{equation}
Recall that geometrically the first functional is associated with a tetrahedron ($\ell$ and $\tilde{\ell}$ being a pair of opposite edges) and the second one to a quadrilateral (actually two triangles hinged by one of its diagonal, $\ell$ or $\tilde{\ell}$) bounding, so to speak,  a portion of volume of amount $\lambda_k$, the eigenvalue of the volume operator given in (\ref{kbasisl}).
In order to select in a convenient way the Hilbert space on which the volume operator acts and all the functionals above can be defined consistently, the role of Regge symmetries, originally introduced for the $6j$ \cite{Regge59}, is crucial.  Such  symmetries in their original formulation are recognized as functional relations on the arguments (namely they cannot be derived by interchanging  the $6j$ arguments as happens for the so--called `classical' or tetrahedral symmetries)  and read
\begin{equation}\label{Rsym1}
\begin{Bmatrix}
a & b & \ell \\
c & d & \tilde{\ell} 
\end{Bmatrix}\,=\,
\begin{Bmatrix}
s-a & s-b & \ell \\
s-c & s-d & \tilde{\ell} 
\end{Bmatrix}\,:=\,
\begin{Bmatrix}
a' & b' & \ell \\
c' & d' & \tilde{\ell} 
\end{Bmatrix}\,,
\end{equation}
where $s= (a+b+c+d)/2$ is the semi--perimeter of the parameter  quadrilateral
and in the last equality the new  set $(a',b',c',d')$ is defined. 
It can be checked that the total
number of classical and Regge symmetries is 144, which equals  the order of
the product permutation group $S_4 \times S_3\,$.
 
Denoting by $a$ the smallest value among the eight  parameters $(a,b,c,d,a',b',c',d')$,
it can be shown that a consistent ordering of the other parameters
compatible with all the due inequalities is given by 
$\{\,a \leq b \leq d\; ;\; d-(b-a) \leq c \leq d+ (b-a)\,\}\,$.
This sort of  gauge fixing  implies that the whole problem becomes finite--dimensional
and workable out for each fixed values of the parameters $(a,b,c,d)$ $\in \mathbb{R}^4$. Moreover:
{\em i)} the tetrahedron $<\tilde{\ell}\,|\,\ell\,>\,(a,b,c,d)$ can be chosen as the
reference one, calling  $<\tilde{\ell}\,|\,\ell\,>\,(a',b',c',d')$ its Regge--conjugate;
{\em ii)} the same thing holds for the quadrilateral denoted $ <\ell\,|\,k\,>\,(a,b,c,d)$
and its conjugate $<\ell\,|\,k\,>\,$ $(a',b',c',d')$. 
More technical details about this specific  parametrization and the denomination  Regge--`conjugate' (as well as the proof that  the volume operators and all quantities in its three--term recursion relation (\ref{SymRR}) are Regge--invariant) can be found in \cite{Aquila0} and 
\cite{AqMaMaJPA2013} respectively. 

Coming back to the statement regarding the correspondence (\ref{dict2}), the remarks above should have made clear that Regge symmetry is strictly related to the duality property of the Racah algebra discussed at the end of section \ref{algebra}. 
Note that in \cite{GrZhSovPh1988} it had been already recognized that (classical + Regge) symmetries do have the group structure given by $S_4 \times S_3$, to be identified with the automorphism group of the Racah algebra.

\section{Classification of discrete polynomial families}\label{class}

In this section the focus will be on interconnections among the families of discrete orthogonal polynomials in view of the formalization presented in section  
\ref{algebra} and summarized there in items {\bf i)} and  {\bf  ii)}. This analysis --not addressed elsewhere to our knowledge-- is just sketched here, leaving aside a number of technical details that can be found in \cite{Tesi}. 
The various cases, together with the most significant properties of each family, are summarized in the following table.\\

\begin{center}
{\bf Finite families of discrete orthogonal polynomials  [\,$(a,b,c,d)$ fixed\,]}
\end{center}

\begin{center}
\begin{tabular}{||c|c||c|c|c|c||}
\# & family & orthogonality on lattice &  eigenvalue & degree \\ 
\, &\,  & \, & \small{(related to the  variable)}  & related to \\
\hline \hline
\, &\,  & \, & \,  &\,\\  
 {\bf I.A} & $<\tilde{\ell}\,|\,\ell\,>$ & 
$\sum_{\tilde{\ell }} \,\overline{<\tilde{\ell}\,|\,\ell '\,>}<\tilde{\ell}\,|\,\ell\,> = 
\delta_{\ell ' \ell }$
&  $\ell(\ell +1)$ & $\tilde{\ell}$\\ 
\, &\,  & \, & \, &\, \\  
\hline
\, & \,  & \, & \, & \, \\ 
{\bf I.B} & $<\ell\,|\,\tilde{\ell}\,>$ & $\sum_{\ell}\, 
\overline{<\ell\,|\,\tilde{\ell} '\,>}
<\ell\,|\,\tilde{\ell}\,> = 
\delta_{\tilde{\ell}' \tilde{\ell}}$ &  $\tilde{\ell}(\tilde{\ell} +1)$ & $\ell$ \\
\, & \,  & \, & \, &\, \\  
\hline
\hline
\, & \,  & \, & \, & \, \\  
{\bf II.A}  & $<\ell\,|\,k\,>$ & $\sum_{\ell} \,\overline{<\ell  \,|\,k' \,>} 
<\ell\,|k\,> = 
\delta_{k' k }$ & $\lambda_k$  & $\ell$ \\
\, & \,  & \, & \, &\, \\   
\hline
\,  & \, & \, &\, \, &\\  
{\bf II.B}  & $<k\,|\ell\,>$ & $\sum_{k} \,\overline{<k \,|\,\ell'\,>} <k\,|\ell\,> = 
\delta_{\ell' \ell }$ & $\ell(\ell +1)$  & $k$ \\ 
\, & \,  & \, & \, &\, \\  
\hline
\hline
\, & \,  & \, & \, &\, \\  
{\bf III.A} & $<\tilde{\ell }\,|\,k\,>$ & $\sum_{\tilde{\ell}} \,\overline{<\tilde{\ell}  \,|\,k' \,>} <\tilde{\ell}\,|k\,> = 
\delta_{k' k }$ & $\lambda_k$  & $\tilde{\ell}$ \\
\, & \,  & \, & \, &\, \\   
\hline
\, & \,  & \, & \, &\, \\  
{\bf III.B}  & $<k\,|\tilde{\ell}\,>$ & $\sum_{k} \,\overline{<k \,|\,\tilde{\ell}'\,>} <k\,|\tilde{\ell}\,> = 
\delta_{\tilde{\ell}' \tilde{\ell} }$ & $\tilde{\ell}(\tilde{\ell} +1)$  & $k$ \\ 
\, & \,  & \, & \, &\, \\  
\hline
\hline
\end{tabular}
\end{center}
Comparing the notations adopted here --the bar stands for complex conjugation or simply transposition in the real cases-- with  those of section \ref{algebra}, it is straightforward to recognized that the classes {\bf I}, 
{\bf II} and {\bf III}  are in correspondence with the overlap functions in (\ref{over1e2}), 
(\ref{over1e3}) and (\ref{over2e3}) (restricted to finite sets  by suitable choices of the omitted parameters), respectively. 

Looking at the family {\bf IA}, observe that $\overline{<\tilde{\ell}\,|\,\ell '\,>}$ $:=$
$<\ell '\,|\,\tilde{\ell}\,>$ $=$ $<\tilde{\ell}\,|\,\ell '\,>$ by the convention chosen for $6j$ symbols in (\ref{seij}) (and similarly for  {\bf IB}). 
Thus  `self--duality' relations for  class {\bf I} read either way
\begin{equation}\label{dualityI}
\sum_{\tilde{\ell }} \,<\ell '\,|\tilde{\ell}\,\,><\tilde{\ell}\,|\,\ell\,> \,= \,
\delta_{\ell ' \, \ell }\;\; \;\; \text{and} \;\; \;\;
\sum_{\ell}\, 
<\,\tilde{\ell} '\,|\,\ell \,>
<\ell\,|\,\tilde{\ell}\,> \,= \,
\delta_{\tilde{\ell}' \, \tilde{\ell}}\,\,,
\end{equation}
once fulfilled the completeness relations
$\Sigma \,|\,\tilde{\ell} \, ><\,\tilde{\ell}\,|$ $= \mathbb{I}$ and
$\Sigma \,|\,\ell \, ><\,\ell\,|$ $= \mathbb{I}$
for the binary coupled eigenbases introduced in (\ref{jbasis}).
Note  that the operators associated with class {\bf I} ($\mathbf{J}_{12}^2$ and
$\mathbf{J}_{23}^2$) represent  a `Leonard pair'
so that the associated overlap functions (recoupling coefficients) are 
necessarily hypergeometric of Racah type \cite{Leo1982}. 
More generally, in connection with  the analysis of the other classes, 
a stringent result holds true:  any {\em finite} system of orthogonal polynomials  whose dual is a finite system of orthogonal polynomials must be 
(possibly $q$--deformed) Racah or one of its limiting cases which constitute finite systems 
(refer to \cite{Koek2010} for a modern monograph
on hypergeometric polynomials in the Askey--Wilson scheme). 
Indeed here all of the families 
are consistently defined, for fixed parameters $(a,b,c,d)$, as finite sets 
(recall the choice on the ordering discussed in connection with Regge symmetry)
but the recognition of classes  {\bf II} and {\bf III} as belonging to the Askey scheme
is certainly not straightforward. (More precisely, the reduction process to specific hypergeometric functions of 
type $_4F_3$ would require to find out a closed algebraic form for the sets of eigenvalues of the volume operator for given parameters, a task  not yet accomplished.)

For what concerns duality within  class 
{\bf II}, a first remark is about the bar operation:  
$\overline{<\ell  \,|\,k \,>}$ is $<\,k\,| \ell \,>$, but the latter, unlike 
what happens for the $6j$, is not necessarily equal to $<\ell \,|\,k \,>$ because 
this property actually depends on the volume operator $K$
being Hermitian (imaginary antisymmetric) \cite{CaCaMaCQG2002} or 
real symmetric (see \cite{AqMaMaJPA2013}
also for plots of the family of eigenfunctions $<\ell  \,|\,k \,>$).
Anyway, both options can be included through a suitable notation into the duality relations 
\begin{equation}\label{dualityII}
\sum_{\ell} \,<\,k'\,|\ell\,><\,\ell\,|\,k\,> \,= \,
\delta_{ k' \, k }\;\; \;\; \text{and} \;\; \;\;
\sum_{k}\, 
<\,\ell'\,|\,k \,>
<\,k\,|\,\ell\,> \,= \,\pm 
\delta_{\ell' \, \ell}
\end{equation}
 according to the choiche of the representation of $K$.
Duality relations in class {\bf III} are similar to 
(\ref{dualityII}), with $\tilde{\ell}$ taking the role  of $\ell$.

To conclude this general overview on duality relationships, a further remarkable property --transversal with respect to the classes-- has to be mentioned, namely   
\begin{equation}\label{dualityIII}
\sum_{k} \,<\,\tilde{\ell} |\,k>\,<\,k\,|\,\ell\,> \,= \,
 \,= \,\pm \,
<\tilde{\ell}\,|\,\ell\,>\,.
\end{equation}
Such  a `triangular relation' (and the other ones that can be derived by using 
the properties of the single classes given above) closely resembles the Racah identity
satisfied by three $6j$ symbols and might be used also to explore
a formalization of the whole subject within the general scheme of tensor categories.

\section{Limiting cases}\label{limiti}
The issue of asymptotic (semiclassical) limits of angular momentum functions
is of continuous interest in many fields, ranging from special function theory
\cite{Koek2010} to applied quantum mechanics \cite{Aquila1}. 
Here just a few remarks concerning two limiting cases of  families {\bf II.A} 
and {\bf III.B} are sketched.
 
The reference model of asymptotics  is the well--know limit of the $6j$ symbol for three large entries 
(see \cite{PR,Russi}),
$6j$ $\rightarrow$ $3j$, where the latter is the Wigner symbol,  the symmetrized version of a Clebsch--Gordan coefficient. The counterpart of this operation in the Askey scheme is achieved by moving one step downwards from  top, namely from
$_4F_3$ (Racah) to $_3F_2$ (Hahn and dual Hahn) hypergeometric families.
  
A new change of notation is needed which consists in restoring the string 
$(j_1,j_2,j_3,j_4)$ for the parameters (see (\ref{Conv2})) and in writing down
as an array  the functions in (\ref{Conv3})  (equivalently, in family  {\bf II.A})  
according to 
\begin{equation}\label{Conv4}
\Psi_{\,\ell}^{(k)}(j_1,j_2,j_3,j_4) \,\; \rightarrow \,\,\,
\begin{Bmatrix}
j_1 & j_2 & \vert & \ell \\
j_3 & j_4 & \vert & \lambda_k
\end{Bmatrix},
\end{equation}
where the vertical bars in front of the last column of this symbol indicate that not all of the entries are constrained by  standard triangular inequalities, as happens for the $6j$.  
To address any limit in which (some of) the arguments of the symbols become large --a fact that implies that all of the arguments can be `running'-- a convenient notation is to substitute capital to small letters. Thus the formal limiting process for the symbol in 
(\ref{Conv4}) when the arguments of the lower row become large can be displayed as
a generalized $3j$ coefficient, denoted $3\mathfrak{j}$, related in turn to a generalized dual Hahn polynomial;
schematically 
\begin{equation}\label{limit1}
\begin{Bmatrix}
j_1 & j_2 & \vert & \ell \\
J_3 & J_4 & \vert & \Lambda_k
\end{Bmatrix}\;
\,\rightarrowtail \;
\begin{pmatrix}
j_1 & j_2 & \vert & \ell \\
J_4-\Lambda_k & \Lambda_k -J_3 & \vert & J_3-J_4
\end{pmatrix} \; \leftrightarrow \; 3\mathfrak{j}\, 
\;\,(\text{dual Hahn family})\,.
\end{equation}
On applying a similar procedure to family {\bf III.B}, and denoting $\tilde{L}$
the previous generic argument $\tilde{\ell}$ (playing the role of $j_{23}$), 
the resulting correspondence would read 
\begin{equation}\label{limit2}
\begin{Bmatrix}
j_1 & j_2 & \vert & \lambda_k \\
J_3 & J_4 & \vert & \tilde{L}
\end{Bmatrix}\;
\;\rightarrowtail \;\;
\begin{pmatrix}
j_1 & j_2 & \vert & \lambda_k \\
J_4-\tilde{L} & \tilde{L} -J_3 & \vert & J_3-J_4
\end{pmatrix} \leftrightarrow \; 3\mathfrak{j} \,
\;\,( \text{Hahn family})\,.
\end{equation}
A few  comments on these results are in order, leaving aside a more careful analysis
and most technical details  reported in \cite{Tesi}. As already noticed, the symbols 
in round brackets on the right--hand sides of 
(\ref{limit1}) and (\ref{limit2}) are generalized counterparts of $3j$ coefficients,
the arguments in the lower row being interpreted as magnetic quantum numbers. 
They actually share with standard $3j$s a suitable formulation of  Regge symmetry \cite{Regge58}
and their properties as orthogonal families are inferred from three--term recursion relationships. The latter can in turn be derived  as limits of  the three--term recursions
at the upper level (in particular, the relation for (\ref{limit1}) can be quite 
easily worked out). The motivation for associating dual Hahn and Hahn families respectively is related with the specific lattices these three--term recursion relations are 
defined on. Thus it is found that the relation for (\ref{limit1}) mimics  the behavior 
of the relation of a $3j$ on a quadratic lattice ($\ell (\ell +1)$), so that it is functionally similar to the standard dual Hahn polynomial family. Conversely,
 the relation for (\ref{limit2}) mimics  the behavior 
of the relation of a $3j$ on a linear lattice (given by scaling the quantum number 
$m \equiv J_3-J_4$) and thus these functions represent  counterparts of the
Hahn polynomial family.


\section{Outlook}\label{concl}
Further developments can be addressed in parallel, from algebraic--analytical and geometric
viewpoints. A schematic list of ongoing works (and still open questions) follows:
\begin{itemize}
\item improved interpretations of Regge symmetries on the geometric ({\em scissor--congruent tetrahedra} \cite{Rob1999}) and algebraic ({\em quaternionic reparametrization} \cite{Tesi}) sides;
\item convolution rules for overlap functions (specifically, symmetric recoupling coefficients) of Racah algebra;
\item composition rules of collections of quadrilaterals able to provide new classes of integrable quantum systems to be associated with
extended quantum geometries;
\item $q$--deformed extensions and limiting cases of the dual sets of orthogonal polynomials also in view of applications in quantum chemistry. 
\end{itemize}
In particular, a systematic study of limiting procedures --to be carried out on recurrence relations, on families of polynomials and possibly directly on the defining relations (\ref{CommRel}) of the underlying quadratic algebras-- seems particularly promising  also in view of recent analytical and numerical work on strictly related issues \cite{Aquila2,Aquila3,Aquila4,Aquila5}.

\section*{Acknowledgments}
D M and A M acknowledge partial support from PRIN 2010-2011 {\it Geometrical and analytical theories of finite and infinite dimensional Hamiltonian systems}.


\section*{References}
\begin{thebibliography}{9}

\bibitem{AqMaMaJPA2013}
Aquilanti V, Marinelli D and Marzuoli A 2013 
Hamiltonian dynamics of a quantum of space: hidden symmetries and spectrum of the volume operator, 
and discrete orthogonal polynomials
{\it J. Phys. A: Math. Theor.} {\bf 46} 175303


\bibitem{CaCaMaCQG2002}
Carbone G, Carfora M and Marzuoli A 2002 
Quantum states of elementary three--geometry
{\it Class. Quantum Grav.} {\bf 19} 3761


\bibitem{BiLo9}
Biedenharn L C and  Louck J D 1981
{\it The Racah--Wigner Algebra in Quantum Theory}
(Encyclopedia of Mathematics and its Applications Vol 9)
ed G-C Rota (Reading, MA: Addison--Wesley)

\bibitem{CaMaRa2009}
Carfora M, Marzuoli A and Rasetti M
2009 Quantum tetrahedra 
{\it J. Phys. Chem. A} {\bf 113} 15376

\bibitem{Regge59}
Regge T 1959 Symmetry properties of Racah's coefficients {\it Nuovo Cimento}  {\bf 11} 116

\bibitem{GrLuZhAOP1992}
Granovskii Ya I, Lutzenko I M and Zhedanov A S 1992
Mutual integrability, quadratic algebras, and dynamical symmetry
{\it Ann. Phys.} {\bf 217} 1

\bibitem{GrZhSovPh1988}
Granovskii Ya I and  Zhedanov A S 1988
Nature of the symmetry group of the 6j-symbol
{\it Sov. Phys. JETP} {\bf 67} 1982

\bibitem{Skly1982}
Sklyanin E K 1982 Some algebraic structures connected with the Yang--Baxter equation
{\it Funct. Anal.} {\bf 16} 263

\bibitem{Askey}
Askey R 1975 {\it Ortogonal Polynomials and Special Functions}
(Philadelphia, PE: SIAM)


\bibitem{Russi}
Varshalovich  D A,   Moskalev A N and  Khersonskii V K 1988
{\it Quantum Theory of Angular Momentum}
(Singapore: World Scientific)


\bibitem{Aquila0}
Aquilanti V, Haggard H M, Hedeman A, Jeevanjee N, Littlejohn R G and Liang Y  2012
Semiclassical mechanics of the Wigner 6j-symbol
{\it J. Phys. A: Math. Theor.} {\bf 45} 065209


\bibitem{Tesi}
Marinelli D 2013
{\it Single and Collective Dynamics of Discretized Geometries}
PhD Thesis  (Pavia: University Press) 

\bibitem{Leo1982}
Leonard D A 1982
Orthogonal polynomials, duality and association schemes
{\it SIAM J. Math. Analys.} {\bf 13} 656

\bibitem{Koek2010}
Koekoek R, Lesky P A and Swarttouw  R F 2010
{\it Hypergeometric Orthogonal Polynomials and Their q-Analogues}
(Heidelberg Dordrecht London New York: Springer)



\bibitem{Aquila1}
Ragni M, Bitencourt A C P, Da S Ferreira C, Aquilanti V, Anderson R W and Littlejohn R G 2009
Exact computation and asymptotic approximations of 6j symbols: illustration of their semiclassical limits
{\it Int. J. Quantum Chem.}
{\bf 110} 731


\bibitem{PR}
Ponzano G and Regge T 1968
{\it Semiclassical limit of Racah coefficients} in
{\it Spectroscopic and Group Theoretical 
Methods in Physics}  ed F Bloch et al 
 (Amsterdam: North--Holland) p 1
 
 
 \bibitem{Regge58}
Regge T 1958 Symmetry properties of Clebsch--Gordan's coefficients 
{\it Nuovo Cimento}  {\bf 10} 544

\bibitem{Rob1999}
Roberts J 1999 Classical 6j-symbol and the tetrahedron {\it Geom. Topology}  {\bf 3} 21




\bibitem{Aquila2}
 Bitencourt A C P, Marzuoli A, Ragni M, Anderson R W and Aquilanti V 2012
 Exact and asymptotic computations of elementary spin networks: classification of the quantum--classical boundaries
 {\it Lect. Notes Comput. Science} {\bf 7333} Part I
ed  B Murgante et al 
 (Berlin Heidelberg: Springer--Verlag) p 723


\bibitem{Aquila3}
Anderson R W, Aquilanti V, Bitencourt A C P, Marinelli D and Ragni M  2013
The screen representation of spin networks: 2D recurrence, eigenvalue equation for 6j-symbols, geometric
interpretation and Hamiltonian dynamics
  {\it Lect. Notes Comput. Science} {\bf 7972}
 (Berlin Heidelberg: Springer--Verlag) p 46
 
\bibitem{Aquila4}
Ragni M, Littlejohn R G, Bitencourt A C P, Aquilanti V and Anderson R W  2013
The screen representation of spin networks: images of 6j-symbols and semiclassical features 
  {\it Lect. Notes Comput. Science} {\bf 7972}
 (Berlin Heidelberg: Springer--Verlag) p 60
 
 \bibitem{Aquila5}
Calderini D, Coletti C and Aquilanti V   2013
Continuous and discrete algorithms in quantum chemistry: polynomial sets, spin networks 
and Strumian orbitals
  {\it Lect. Notes Comput. Science} {\bf 7972}
 (Berlin Heidelberg: Springer--Verlag) p 32
 





\end{thebibliography}
\end{document}